\documentclass[aps,pra,superscriptaddress,amsmath,amssymb,twocolumn,amsfonts,floatfix, longbibliography]{revtex4-2}
\usepackage{graphicx}
\usepackage{epstopdf}
\usepackage{amsbsy}
\usepackage{bm}
\usepackage{mathrsfs}

\usepackage[dvipsnames]{xcolor}
\definecolor{deepfuchsia}{rgb}{0.76, 0.33, 0.76}
\definecolor{electricpurple}{rgb}{0.75, 0.0, 1.0}
\usepackage[colorlinks=true,linktoc=page,linkcolor=blue,citecolor=magenta,urlcolor=electricpurple]{hyperref}
\usepackage{orcidlink}
\usepackage[normalem]{ulem}

\newcommand{\beq}{\begin{equation}}
\newcommand{\eeq}{\end{equation}}
\newcommand{\bea}{\begin{eqnarray}}
\newcommand{\eea}{\end{eqnarray}}
\newcommand{\non}{\nonumber}
\newcommand{\br} { \bm{r} }

\newcommand{\eqn}[1] {Eq.~(\ref{#1})}

\newcommand{\equref}[1]{Eq.~(\ref{#1})}
\newcommand{\equsref}[2]{Eqs.~(\ref{#1}) and (\ref{#2})}

\newcommand{\figref}[1]{Fig.~\ref{#1}}

\newcommand{\appref}[1]{Appendix~\ref{#1}}

\renewcommand{\approx}{\simeq}

\begin{document}


\title{Spectral Bifurcation and Anomalous Supercurrent in Dissipative Topological Insulator-based Josephson Junctions}

\author{{Ardamon Sten}\,\orcidlink{0009-0006-2227-0695}}
\email[]{ardamons21@iitk.ac.in}
\affiliation{Department of Physics, Indian Institute of Technology, Kanpur 208016, India}
\author{{Paramita Dutta}\,\orcidlink{0000-0003-2590-6231}}
\thanks{Jointly supervised this work}
\email[]{paramita@prl.res.in}
\affiliation{Theoretical Physics Division, Physical Research Laboratory, Navrangpura, Ahmedabad-380009, India}
\author{{Sudeep Kumar Ghosh}\,\orcidlink{0000-0002-3646-0629}}
\thanks{Jointly supervised this work}
\email[]{skghosh@iitk.ac.in}
\affiliation{Department of Physics, Indian Institute of Technology, Kanpur 208016, India}

\date{\today}

\begin{abstract}
The interplay between topological protection and dissipation constitutes a critical frontier in the realization of hybrid quantum devices. Here, we investigate the transport signatures in a dissipative topological insulator-based Josephson junction, a platform that directly probes the competition between quantum coherence and loss. We model dissipation by coupling a `lossy' metallic lead to the junction, described effectively by a non-Hermitian Hamiltonian derived using the Lindblad formalism. We observe that the junction exhibits an asymmetric complex Andreev spectrum, where the imaginary energy component imposes a finite lifetime on the quasi-bound states. Furthermore, beyond specific phase intervals, the real component of the spectrum bifurcates: one branch merges with the continuum, while the other penetrates just below the superconducting gap. Crucially, the characteristic zero-energy crossing shifts away from $\phi=\pi$ and acquires a non-zero imaginary component; consequently, the associated Majorana bound states acquire a finite lifetime, signaling a loss of robustness against dissipation. Finally, this spectral asymmetry drives an anomalous supercurrent, manifested as a non-vanishing current at zero phase difference. Our results reveal how dissipation fundamentally reshapes superconducting transport in topological junctions, opening new directions for dissipation-engineered quantum devices.

\end{abstract}

\maketitle


Josephson junctions (JJs) play a pivotal role in superconducting electronics, offering a versatile tool for probing quantum transport, topological states, and unique superconducting behaviors~\cite{Wolf2017}. In conventional Hermitian JJs, the superconducting phase difference determines the current-phase relationship (CPR), which is typically $2\pi$-periodic and sinusoidal in `short' junctions~\cite{Josephson1974}. These junctions are fundamental to technologies such as Superconducting Quantum Interference Devices (SQUIDs) and single-photon detectors~\cite{walsh2021}. Topological insulator-based Josephson junctions (TIJJs) provide an ideal setting for realizing exotic quantum states~\cite{Fu2008,Cayao2022a,Dutta2023,Dutta2024}, where proximity-induced superconductivity on the surface of a 3D topological insulator (3DTI) induces helical Andreev bound states (ABSs)~\cite{Fu2008,Tkachov2013,Olund2012}. These junctions support non-chiral Majorana modes~\cite{Fu2008,Hasan2010,Qi2011}, leading to $4\pi$-periodic Josephson effects~\cite{Fu2008,Fu2009,Olund2012,wiedenmann2016}.

Dissipation is an intrinsic aspect of real-world Josephson junctions, where non-Hermitian Hamiltonians capture the effects of environmental coupling~\cite{cayao2023non,Kornich2023,Li2024,Shen2024,cayao2024non,Beenakker2024}. In non-Hermitian Josephson junctions (NH-JJs), dissipation generates complex spectra and exceptional points (EPs), producing behavior absent in Hermitian systems~\cite{Li2024,cayao2023non,Shen2024,Pino2025}. Although EPs in the ABS spectrum may leave no clear imprint on static quantities such as the DC Josephson current~\cite{Shen2024}, they do manifest in dynamical observables like the current susceptibility~\cite{Shen2024} and in thermodynamic measures such as entropy~\cite{Pino2025}. However, the impact of non-Hermiticity on topological Josephson junctions remains largely unexplored, particularly in systems where dissipation competes with topological protection.

\begin{figure}[!b]
    \centering
    \includegraphics[width=0.99\columnwidth,trim={0.0cm 0.1cm 0.2cm 0.1cm},clip]{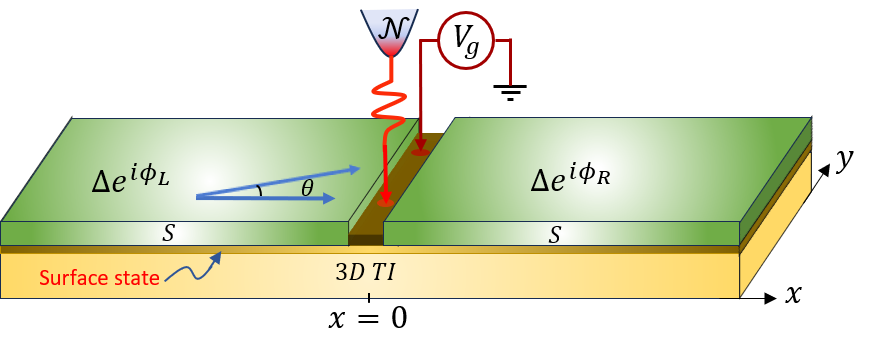}
    \caption{\textbf{Schematic of a non-Hermitian planar TIJJ:} A `short' TIJJ formed by placing two bulk $s$-wave singlet superconductors on the surface of a 3DTI. A top gate and a `lossy' metallic lead ($\mathcal{N}$) are attached at the junction ($x=0$), where the gate modifies the potential barrier via an applied voltage $V_g$ and the lead introduces dissipation due to coupling to an electron reservoir. }
    \label{fig:junction_schematic}
\end{figure}

In this work, we address the central question: how does dissipation caused by coupling a `lossy' metallic lead~\cite{cayao2023non,Ohnmacht2024} at the junction, affect the quantum transport properties of TIJJs? To explore this, we model the dissipative effects of the lead by an effective non-Hermitian Hamiltonian~\cite{Harrington2022,Li2024} and investigate transport in a non-Hermitian planar Josephson junction on the surface of a 3DTI (NH-TIJJs), as illustrated schematically in \figref{fig:junction_schematic}. Here, proximity-induced superconductivity intertwines with non-Hermitian effects, offering a unique platform to study the interplay between dissipation and topological superconductivity.

\noindent \textit{Effective model of a NH-TIJJ:} The NH-TIJJ in \figref{fig:junction_schematic} consists of two $s$-wave singlet superconductors with identical gap $\Delta$ and phases $\phi_L$ and $\phi_R$ on the left and right, respectively. Dissipation due to the `lossy' lead~\cite{cayao2023non,Ohnmacht2024} connected to an electron reservoir is modeled in a minimal way, within the Lindblad formalism neglecting quantum jumps~\cite{breuer2002,daley2014} (see \appref{Lindblad_formalism} for details), by an imaginary potential barrier at the junction~\cite{Li2024}. The combined effect of the top gate and the lead is therefore captured by a complex contact junction potential barrier: $U(x) = (V_1 - iV_2)\delta(x)$, where $V_1 > 0$ represents the real barrier strength tunable by the applied gate voltage $V_g$ and $V_2 > 0$ accounts for the imaginary barrier strength induced by the lead.

\begin{figure*}[ht]
\centering
\includegraphics[width=\textwidth]{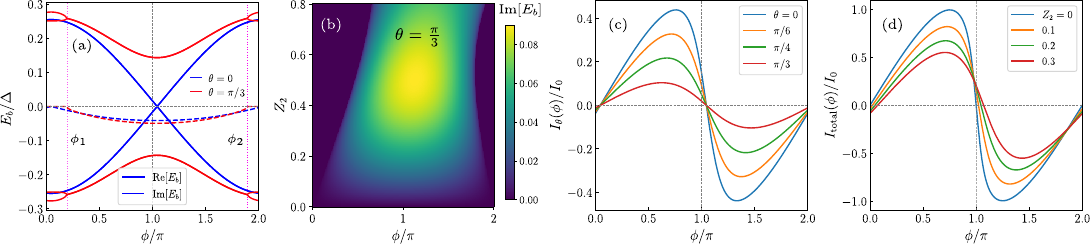}
\caption{\textbf{Characteristics of the NH-TIJJ for a complex barrier ($Z_1\ne 0$, $Z_2 \ne 0$) in the tunneling limit ($\Gamma_0\ll \Delta$):} (a) The real (solid lines) and imaginary (dashed lines) parts of the Andreev spectrum of the NH-TIJJ as a function of the phase difference $\phi$ plotted for different angles of incidence $\theta$ taking the imaginary barrier parameter $Z_2=0.2$. The $\phi_1$ and $\phi_2$ denote the GEPs, beyond which the real part of the spectrum exits the gap and the imaginary part of the spectrum vanishes. (b) Variation of the imaginary part of the spectrum as a function of $\phi$ and $Z_2$ for the incident angle $\theta=\pi/3$, the boundary of the dark purple region demonstrates the asymmetric gap-exit points in the spectrum. (c) The zero temperature angle-resolved current $I_{\theta}(\phi)$ plotted as a function of $\phi$ for different angles of incidence. (d) The total supercurrent $I_{\mathrm{total}}(\phi)$ as a function of $\phi$ for different barrier parameters. We have chosen the real barrier parameter $Z_1=0.4$, tunneling parameter $\gamma=0.3$ and $I_0=eN\Gamma_0/(2\hbar)$ with $N$ being the number of open channels.}
\label{NHTIJJ_complex}
\end{figure*}

Assuming tunnel coupling between the superconductors and the 3DTI surface state~\cite{Tkachov2013,Fu2008,Potter2011,Lababidi2011}, the physics of the proximity-induced superconductivity on the surface state can be described by the effective Bogoliubov-de Gennes (BdG) Hamiltonian~\cite{Tkachov2013,Fu2008}: 
\bea
\mathcal{H}_{\rm eff} & = & \Phi^{\dagger} H_{\rm BdG} \Phi\;,
\non\\
H_{\rm BdG} & = & \begin{bmatrix}
    \frac{\hbar}{i} v_F \bm{\sigma}\cdot \nabla - \mu - \Gamma(\epsilon) & \Tilde{\Delta}(\epsilon) i \sigma_y e^{i \phi_{L,R}} \vspace{0.1cm}\\ 
    - \Tilde{\Delta}(\epsilon) i \sigma_y e^{-i \phi_{L,R}} & \frac{\hbar}{i} v_F \bm{\sigma}^*\cdot \nabla + \mu - \Gamma(\epsilon)
\end{bmatrix} , \non \\
\Tilde{\Delta}(\epsilon) &=& \frac{i\Gamma_0 \Delta}{\sqrt{\epsilon^2-\Delta^2}}\;,\Gamma(\epsilon)=\frac{i\Gamma_0\epsilon}{\sqrt{\epsilon^2-\Delta^2}}\;, \Gamma_0 = \pi t^2 \mathcal{D}_0;
\eea
written in the Nambu basis $\Phi^{\dagger} = (\varphi^{\dagger}_{\uparrow},\varphi^{\dagger}_{\downarrow},\varphi_{\uparrow},\varphi_{\downarrow})$ with $\varphi^{\dagger}_{s}$ being the electron creation operator with spin $s = \uparrow, \downarrow$. $\mu$ is the chemical potential of the superconductor, $v_F$ is the Fermi velocity of the surface states of the 3DTI and $\sigma_i$ ($i=1,2$) are the Pauli matrices in spin space. $\Tilde{\Delta}(\epsilon)$ is the proximity induced superconducting pairing potential which is a function of the Bogoliubov quasi-particle energy $\epsilon$, $\Gamma(\epsilon)$ is the shift in energy of the quasi-particles due to tunneling with amplitude $t$, and $\Gamma_0$ is the tunneling energy scale that determines the tunneling rate. $\mathcal{D}_0$ is the density of states of the superconductors in the normal state. The proximitized superconductivity on the 3DTI surface is determined by the tunneling energy $\Gamma_0$, in the tunneling regime ($\Gamma_0\ll\Delta$) the induced gap is much smaller than the parent s-wave gap and in the strong proximity limit ($\Gamma_0=\Delta$) the pairing potential is equal to that of the parent superconductor. In this work, we investigate the transport properties in both of these regimes to see whether any new feature is observed in the tunneling regime.

To investigate transport in the ``short'' NH-TIJJ (\figref{fig:junction_schematic}), assuming the width of the junction much smaller than the coherence length of the superconductors, we consider $\phi_L=0$ and $\phi_R=\phi$ without loss of generality. In this limit, the physics of the system can be described by~\cite{Tkachov2013}:
\beq
\left[H_{\text{BdG}}(\br) + (V_1\tau_z -i V_2 \tau_0)\sigma_0 \right]\Psi(\br)=\epsilon\Psi(\br)\;,
\label{BdG_eq}
\eeq
where, $\tau_z$ is the third Pauli matrix in the Nambu space, $\sigma_0$ is the $2 \times 2$ identity matrix and $\Psi(\br) = \left[\psi_{\uparrow,\epsilon} (\br),\psi_{\downarrow,\epsilon}(\br),\psi^*_{\uparrow,-\epsilon}(\br),\psi^*_{\downarrow,-\epsilon}(\br)\right]^T$ is the Nambu spinor wavefunction. Quasiparticles enter the barrier from the left at an angle of incidence $\theta \in [-\pi/2, \pi/2]$ and we assume translational invariance along the $y$-direction. Klein tunneling is insensitive to the precise form of the barrier~\cite{Tkachov2013,Tkachov2013a}, so the specific shape of the potential does not matter for our analysis; only the influence of non-Hermiticity is relevant.

\noindent \textit{Complex helical ABSs and CPRs:} The properties of the ABSs in the NH-TIJJ can be obtained by solving \equref{BdG_eq}. Since wavefunctions of the ABSs decay exponentially as $|x|\rightarrow \infty$, with translational invariance in the $y$-direction, the wavefunctions have the general form: $\Psi(\br)=\sum_{k_y}\psi_{k_y}(x)e^{ik_yy}$ (see Appendix \ref{wfn_NHTIJJ} for explicit forms of the ABS wavefunctions within the Andreev approximation). Integrating \equref{BdG_eq} around $x=0$ and defining $Z=Z_1-iZ_2$ where $Z_{1,2}=V_{1,2}/(\hbar v_F)$ characterize the potential barrier strengths, we get the boundary condition,
\begin{equation}\label{boundary}
    \psi_{k_y}(0_-)=[1+iZ_1\tau_z \sigma_x+Z_2 \tau_0 \sigma_x]\psi_{k_y}(0_+)\;,
\end{equation}
which leads to an equation in terms of $ \mathcal{E}/\tilde{\Delta}$ (see \appref{wfn_NHTIJJ}):
\begin{equation}\label{gen_ABS_eq}
    -4iZ_2 \cos{(\theta)}\frac{\mathcal{E}}{\tilde{\Delta}}\sqrt{1-\left(\frac{\mathcal{E}}{\tilde{\Delta}}\right)^2}= \mathcal{F(\theta,\phi)}-2T_2\left(\frac{\mathcal{E}}{\tilde{\Delta}}\right)^2 \,.
\end{equation}
Here, $\mathcal{F}(\theta, \phi)= (1+Z^2_1-Z^2_2)\cos^2{(\theta)}\cos{(\phi)}+ (Z^2_1+Z^2_2)(1+\sin^2{(\theta)})+\cos^2{(\theta)} + 2 Z_1Z_2\cos^2{(\theta)} \sin{(\phi)}$, $T_2=Z_1^2+Z_2^2+\cos^2{(\theta)}$, $\mathcal{E}(\epsilon)=\epsilon+i\Gamma(\epsilon)$ and $\tilde{\Delta}=\tilde{\Delta}(\epsilon)$. The above equation can be rearranged into a quadratic equation in the variable $y=\left|\frac{\mathcal{E}}{\tilde{\Delta}}\right|^2$ as $Ay^2+By+C=0$ giving the solutions
\beq \label{full_gap_ABS}
\left|\frac{\mathcal{E}}{\tilde{\Delta}}\right|^2=\frac{-B\pm \sqrt{B^2-4AC}}{2A}\;,
\eeq
where $A=4Z^2_2\cos^2{(\theta)}-T^2_2$, $B=\mathcal{F}(\theta, \phi)T_2-4Z^2_2\cos^2{(\theta)}$, $C=-\mathcal{F}^2(\theta, \phi)/4$. In the case of strong proximity effect, $\mathcal{E}$ and ${\tilde{\Delta}}$ are independent of the quasiparticles energy $\epsilon$ and the induced gap can be treated as constant. In this limit, \equref{full_gap_ABS} can be interpreted as the bound state solutions for the TIJJ whose induced gap is equal to that of the parent superconductor (see \appref{wfn_NHTIJJ} for more discussion). Denoting the R.H.S of \equref{full_gap_ABS} as $\mathcal{S}(\theta, \phi)$, we substitute the expressions of $\mathcal{E}$ and $\tilde{\Delta}$ and arrive at the following equation for the bound states in terms of $\alpha=|\frac{\epsilon}{\Delta}|$ in the tunneling regime
\beq \label{ABS_eq}
\alpha^2(1-\alpha^2)=\gamma^2\left(\sqrt{\mathcal{S}(\theta,\phi)}-\alpha\right)^2\;,
\eeq
where $\gamma=\Gamma_0/\Delta\ll1$ is a dimensionless parameter characterizing the tunneling strength. We seek solutions of \equref{ABS_eq} in powers of $\gamma$~\cite{Tkachov2013}: $\left|\epsilon/\Delta \right|=c_1 \gamma+c_2 \gamma^2+c_3 \gamma^3$, where $c_1,c_2$ and $c_3$ are unknown complex coefficients. Then in the `short' junction limit, we obtain two branches (labeled by $\pm$) of complex ABS spectrum with energies (see \appref{wfn_NHTIJJ} for details):
\beq
\frac{E^{\pm}_b(\theta, \phi)}{\Delta}  =  \pm \Bigg[(\gamma-\gamma^2+\gamma^3)\mathcal{S}^\frac{1}{2}(\theta, \phi) + \frac{\gamma^3}{2}\mathcal{S}^\frac{3}{2}(\theta, \phi)\Bigg]\;.
\label{NHTIJJ_ABS_eqn}
\eeq
The real parts of $E^{\pm}_b$ represent physical energies of the ABSs while the negative imaginary part of $E_b^{\pm}$ represents the broadening of  the energy levels due to non-Hermiticity~\cite{Capecelatro2025,Li2024}, the positive imaginary branch signifies pumping of quasiparticles from the reservoir \cite{Kornich2022} and therefore does not represent a physical solution for dissipative systems. The variation of the ABS spectrum for different angles of incidence $\theta$ is shown in \figref{NHTIJJ_complex}(a). Before discussing the features introduced by non-Hermiticity, we first remind ourselves of the ABS characteristics in the Hermitian case; the spectrum for normal incidence ($\theta = 0$) is gapless and features a zero-energy crossing at $\phi=\pi$ which corresponds to two Majorana bound states whose spinors are related by time-reversal symmetry and also preserve Klein tunneling \cite{Tkachov2013}. The spectrum for oblique incidence ($\theta\neq0$), however, features an angle-dependent gap and lacks time-reversal symmetry protection and thus experience scattering by the potential barrier.

In the non-Hermitian case, the normally incident quasiparticles feature a spectrum whose real part is gapless but the zero-energy crossing is shifted away from $\phi=\pi$ in contrast to the Hermitian case. There is also a finite imaginary component accompanying the real counterpart which indicates that Klein tunneling, which exists for normal incidence in the Hermitian case \cite{Tkachov2013}, is not robust against dissipation. The shift of the zero-energy crossing along the $\phi$ axis makes the spectrum asymmetric with respect to $\phi=\pi$ over the interval $\phi \in [0,2\pi]$. Moreover, the finite imaginary component introduces a finite lifetime for the bound states, rendering the associated Majorana bound states unstable in the presence of non-Hermiticity. This can be understood as a purely non-Hermitian effect where dissipation affects the Kramers degeneracy and can induce effects such as splitting of spectral functions of Kramers degenerate states \cite{Deng2021}.

Meanwhile, obliquely incident quasiparticles experience the effect of non-Hermiticity in two qualitatively new effects: (i) broadening of the quasi-ABSs energy levels which means that the bound states have acquired a finite lifetime \cite{datta1997electronic} and (ii) a fully gapped quasi-ABS spectrum which beyond a certain interval $\phi \in [\phi_1, \phi_2]$, the real part of both particle and hole spectrum bifurcates into two branches with one branch going above the induced gap (supra-gap) and the other going below it (sub-gap). We call the points $\phi_1$ and $\phi_2$ on the $\phi$-axis where the real part of the spectrum exits the gap as gap-exit points (GEPs), the imaginary part of the spectrum vanishes beyond these points as shown in \figref{NHTIJJ_complex}(b) and the position of the GEPs, demonstrated by the boundary of the dark purple region, is asymmetric about $\phi=\pi$ in the interval $\phi\in [0,2\pi]$. Since the normally incident quasi-particles do not experience scattering from the real barrier,  therefore the bifurcation is demonstrated only in oblique incidence and is absent for normal incidence.

\begin{figure}[!b]
    \centering
    \includegraphics[width=0.95\columnwidth]{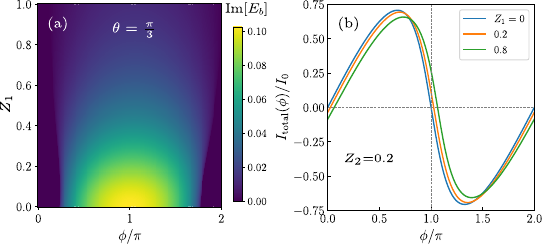}
    \caption{\textbf{Characteristics of the NH-TIJJ for fixed $Z_2 = 0.2$ in the tunneling limit:} Variation of the (a) imaginary part of the Andreev spectrum as a function of $Z_1$ and $\phi$, and (b) total current as a function of $\phi$ for different values of $Z_1$.}
    \label{Z1_variation_fig}
\end{figure}

\begin{figure*}[!ht]
\centering
\includegraphics[width=\textwidth]{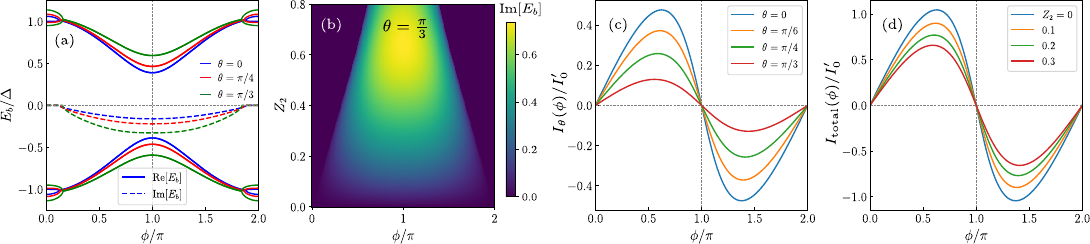}
\caption{\textbf{Characteristics of the ordinary NH-JJ for a complex barrier ($Z_1\neq0, Z_2 \neq  0$):} (a) The real (solid lines) and imaginary (dashed lines) parts of the Andreev spectrum of the 2D NH-JJ as a function of the phase difference $\phi$ plotted for different angles of incidence $\theta$ taking the imaginary barrier parameter $Z_2=0.2$. (b) Variation of the imaginary part of the spectrum as a function of $\phi$ and $Z_2$ for the incident angle $\theta=\pi/3$, the boundary of the dark purple region demonstrates the symmetric gap-exit points in the spectrum. (c) The angle-resolved current $I_{\theta}(\phi)$ plotted as a function of $\phi$ for different angles of incidence at zero temperature and $Z_2=0.2$. (d) The total supercurrent $I_{\mathrm{total}}(\phi)$ as a function of $\phi$ for different barrier parameters. Here, $I'_0=e\Delta/\hbar$ and $Z_1=0.4$. 
}
\label{NHJJ_figure}
\end{figure*}

To calculate the Josephson current, we employ the Furusaki-Tsukada (FT) formula \cite{Furusaki1991, Lu2018} which provides an alternative approach to compute the supercurrent in terms of the Andreev reflection processes as,
\begin{equation}
    I_{\theta}(\phi)=\frac{e}{\hbar}k_BT\sum_{\omega_n}\frac{\tilde{\Delta}(\omega_n)}{\Omega_n}[r_{he}(\phi,\omega_n)-r_{eh}(\phi,\omega_n)]\;,
    \label{FT_formula}
\end{equation}
where analytic continuation for the energy is done as $E\to i\omega_n$ in terms of the Matsubara frequencies $\omega_n=(2n+1)\pi/k_BT$ with $n$ being an integer, the parameter $\Omega_n=\sqrt{\omega^2_n+\Delta^2}$, $r_{he}$ is the Andreev coefficient for an incoming electron-like quasiparticle being reflected as a hole-like quasiparticle and $r_{eh}$ is the corresponding coefficient for the reverse process for a hole-like quasiparticle. In the zero-temperature limit $T\to0$, the summation over Matsubara frequencies can be replaced by an integration over real energy as $k_BT\sum_{\omega_n}\to \frac{1}{2\pi}\int d\omega$ on performing the analytic continuation $i\omega_n\to \omega+i 0^+$. We note that the existing supercurrent formulae for NH systems used to evaluate the current in systems with EPs \cite{Pino2025, Shen2024} incur a divergence of the angle-resolved current at the GEPs, therefore to resolve these divergences we use the FT formula and limit our analysis within the subgap approximation ($|\omega|<\Delta$). The variation of the angle-resolved current $I_{\theta}(\phi)$ for different $\theta$ at zero temperature within the subgap approximation is shown in \figref{NHTIJJ_complex}(c). The current is $2\pi$-periodic and exhibits an `$\textit{anomalous}$' behavior where the supercurrent is non-zero at $\phi=0, 2\pi$ for all angles of incidence. For normal incidence, the current changes sign smoothly and does not exhibit the abrupt discontinuity at the zero-energy crossing as in the Hermitian case, which highlights the contribution of the imaginary part of the spectrum in reshaping and smoothing the CPRs. 
For oblique incidence, in addition to the mentioned characteristics, the critical current decreases with increasing $\theta$ indicating that the effect of non-Hermiticity is more effective with increasing obliqueness.

The total equilibrium Josephson current ($I_{\rm total}$) can be obtained as:
\begin{equation}\label{total_current}
    I_{\text{total}}(\phi) = \int_{-\pi/2}^{\pi/2} I(\phi,\theta)\cos{(\theta)}d\theta \;.
\end{equation}
The CPR for the NH-TIJJ is shown in \figref{NHTIJJ_complex}(d) for different values of the non-Hermiticity strength $Z_2$. The CPR features a $2\pi$-periodic, non-sinusoidal, anomalous current which is asymmetric about $\phi=\pi$ {carrying the signature of the asymmetry of the ABSs shown in Fig.~\ref{NHTIJJ_complex}.} The critical current is maximum for the Hermitian case ($Z_2=0$) and decreases with increasing $Z_2$ for a fixed $Z_1$ due to quasiparticles experiencing decoherence and escaping to the reservoir. The main feature of the CPR is its asymmetry about the phase difference $\phi=\pi$ and its non-zero value at $\phi=0,2\pi$, this anomalous behavior of the supercurrent stems from the asymmetry in the Andreev spectrum whereby the corresponding subgap contribution outside the GEPs is also asymmetric, thus resulting in an anomalous CPR. The asymmetry is an outcome of the interplay between the real barrier and non-Hermiticity, as is evident in \figref{Z1_variation_fig}(a) where the GEPs (boundary of dark purple region) are initially symmetric when $Z_1=0$ and become asymmetric when $Z_1\neq 0$. The variation of the CPR for different values of the real barrier $Z_1$ for a fixed $Z_2$ is shown in \figref{Z1_variation_fig}(b), the critical current is maximum for $Z_1=0$ and decreases as $Z_1$ increases due to enhanced normal reflection. Note that the anomalous behavior is absent when $Z_1=0$ since asymmetry in the spectrum requires both the barrier parameters $Z_1$ and $Z_2$ to be non-zero.

\noindent \textit{Comparison with `ordinary' NH-JJs:} We consider an `ordinary' planar NH-JJ, where the 3DTI in \figref{fig:junction_schematic} is replaced by an `ordinary' metal, to highlight the special features of the NH-TIJJ. Proceeding in a similar way, in the `short' junction limit within the Andreev approximation, the ABS energies are given by (see \appref{NHJJ_wfns} for details),
 \begin{align}
    &g_2(\theta,\phi)\left(\frac{E_b}{\Delta'}\right)^2 = 
    g_1(\theta,\phi) Z^2+\cos^4{(\theta)}\cos^2{\left(\frac{\phi}{2}\right)} \non\\ & + 2Z_1^2\cos^2{(\theta)}\pm 2Z_2\sqrt{g_1(\theta,\phi)\cos^2{\left(\frac{\phi}{2}\right)}-Z^2_1 }.\label{ABS_energy_NHJJ}
 \end{align}
Here, $Z^2=(Z^2_1+Z^2_2)$ with the dimensionless barrier parameters defined as $Z_i=\frac{mV_i}{\hbar^2k_F}$ in this case, $g_1(\theta,\phi) = Z^2-\cos^2{(\theta)}\sin^2{\left(\frac{\phi}{2}\right)}$, $g_2(\theta,\phi) = {\left[Z^2+\cos^2{(\theta)}\right]^2-4Z^2_2\cos^2{(\theta})}$ and $\Delta'$ is the induced superconducting gap amplitude in the metal.


The full Andreev spectrum is shown in \figref{NHJJ_figure} and it shares features similar to that of the NH-TIJJ. The spectrum for different incident angles $\theta$ is shown in \figref{NHJJ_figure}(a). The spectrum is complex with the real branches showing bifurcation beyond the interval $\phi \in[\phi_0,2\pi-\phi_0]$ where $\phi_0(Z, \theta)=2\sin^{-1}\left({\frac{\sqrt{2}Z_2}{\sqrt{\cos^2{(\theta)}+Z^2}}}\right)$, both the energy gap between the real branches and the behavior of the imaginary part are strongly dependent on the angle of incidence. ${\rm Im}[E_b]$ increases with increasing $\theta$ which indicates that the effect of non-Hermiticity increases with increasing $\theta$ consistent with the NH-TIJJ case ({see} \figref{NHTIJJ_complex}). The boundary of the purple region ({see} \figref{NHJJ_figure}(b)) where the imaginary part of the spectrum for $\theta=\pi/3$ is plotted as a function of the imaginary potential strength $Z_2$ and $\theta$ demonstrates the position of the GEPs which are symmetric about $\phi=\pi$ in contrast to the asymmetric feature in NH-TIJJ (see \figref{NHTIJJ_complex}(b)). Hence, the NH-JJ and NH-TIJJ share a common qualitative behavior in that the Andreev spectrum is complex with the real parts exhibiting bifurcation beyond the GEPs while also exhibiting contrasting behaviors such as the absence of zero-energy states in an NH-JJ and an asymmetric spectrum in the NH-TIJJ.

The variation of the angle-resolved Josephson current $I_{\theta}(\phi)$ in \equref{FT_formula} for different angles of incidence is shown in \figref{NHJJ_figure}(c). We note that in stark contrast to the NH-TIJJ, the angle-resolved current does not exhibit the anomalous behavior since the Andreev spectrum of the NH-JJ is symmetric. The total current shown in \figref{NHJJ_figure}(d) for different values of $Z_2$ features a $2\pi$-periodic, non-sinusoidal CPR whose critical current decreases with increasing $Z_2$. The absence of anomalous behavior of the supercurrent in the case of NH-JJ is due to the lack of any Kramers degeneracy in the spectrum, thus there is no interplay with non-Hermiticity to create the asymmetry in the spectrum as in the case of a NH-TIJJ.

\begin{figure}[!b]
    \centering
    \includegraphics[width=0.95\columnwidth]{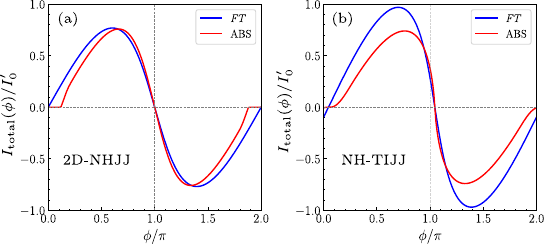}
    \caption{\textbf{Comparison of the CPR obtained from the FT formula (\eqn{FT_formula}) and the ABS formula (\eqn{ABS_CPR}) for a complex barrier ($Z_1=0.4$ and $Z_2=0.2$):} (a) Planar ordinary NH-JJ, and (b) NH-TIJJ in the strong-proximity regime.}
    \label{FT_ABS_cpr}
\end{figure}

\noindent \textit{Comparison with CPR from ABS formula:} The alternative supercurrent formulae for NH systems~\cite{Pino2025, Shen2024} with EPs are not able to resolve the divergence of the supercurrent at the GEPs in our system as mentioned earlier. Therefore we calculate the supercurrent from an ABS formula derived from the polar Green's function \cite{Capecelatro2025} focusing only on the subgap contribution and compare with the CPR computed from the FT formula. The angle-resolved Josephson current at zero temperature from this ABS formula is given by~\cite{Capecelatro2025}:
\bea   
I_{\theta}(\phi)
=& -\frac{2e}{\hbar}\frac{1}{\pi}\Big\{
(\partial_{\phi}\epsilon)
\int_{-\Delta}^{\Delta} d\omega\,
\frac{\tanh(\beta\omega/2)\,\lambda}
{(\omega+\epsilon)^2+\lambda^2}
\non\\
& - (\partial_{\phi}\lambda)
\int_{-\Delta}^{\Delta} d\omega\,
\frac{\tanh(\beta\omega/2)\,(\omega+\epsilon)}
{(\omega+\epsilon)^2+\lambda^2}
\Big\}
\label{ABS_CPR}
\eea
after decomposing the Andreev spectrum as $E_b=\epsilon-i\lambda$ with $\epsilon$ and $\lambda$ being the real and imaginary part of the spectrum respectively. The resulting CPR is plotted in \figref{FT_ABS_cpr} for both the NH-JJ and the NH-TIJJ in the strong-proximity regime. 
Within this subgap approximation, the ABS formula predicts `Josephson gaps', intervals of vanishing supercurrent~\cite{Li2024}. However, a significant discrepancy exists between these two formulas, particularly near the GEPs where the FT result shows no such gaps as shown in \figref{FT_ABS_cpr}. This mismatch arises because the ABS formula captures only the contribution from the bound-state spectrum (Green's function poles), yielding zero current where bound states are absent (beyond GEPs). In non-Hermitian systems, however, a critical contribution stems from the Green's function branch cuts~\cite{Capecelatro2025}. Unlike the ABS formula, the FT approach incorporates all the reflection processes, thereby recovering a finite supercurrent outside the GEPs as well.

\noindent \textit{Summary and conclusions:} We have investigated the quantum transport properties of a dissipative Josephson junction on the surface of a 3DTI, aiming to elucidate the interplay between dissipation and topological protection. Dissipation, modeled via an effective non-Hermitian Hamiltonian within the Lindblad formalism, arises from coupling to a `lossy' metallic lead~\cite{cayao2023non,Ohnmacht2024} at the junction to an electron reservoir. Our study reveals that the Andreev bound state (ABS) spectrum is complex and merges into the continuum regime beyond certain interval of the phase difference. The spectrum also exhibit an asymmetry arising from the interplay of normal scattering and non-Hermiticity. This asymmetry in the spectrum results in the anomalous behavior of the CPR, where a finite supercurrent is present even when the phase difference is zero.


The physics of the `short' NH-TIJJ described in this paper can be realized in practical devices, such as Nb-superconductors on the surface of a HgTe 3DTI~\cite{Maier2012}, with dissipation introduced by attaching a `lossy' metallic lead~\cite{cayao2023non,Ohnmacht2024} at the junction. For such a setup, relevant parameters include $\Delta \approx 1$ meV, $\Gamma_0 \approx 0.2$ meV, $\hbar v_F \approx 250-350$ meV-nm and the junction width $< 1.25 - 1.75$ $\mu$m~\cite{Maier2012}. Another promising platform is Nb-Bi$_2$Te$_3$-Nb JJs~\cite{Stolyarov2022,Charpentier2017}. Since the length of the TIJJ qualitatively changes the CPR~\cite{Lu2023}, a natural future direction is to investigate the effect of dissipation on quantum transport in a `long' NH-TIJJ which will be explored in a separate study. Our findings thus highlight the {effect of dissipation on the Josephson current in a topological Josephson junction}, providing critical insights into the experimental realization of dissipation-engineered hybrid quantum devices and paving the way for leveraging non-Hermitian effects in TIJJs to advance quantum computing architectures.

\subsection*{Acknowledgments}
A.\,S. acknowledges the hospitality of PRL, Ahmedabad, during this work and thanks Amartya Pal and D. Michel Pino for valuable discussions. P.\,D. acknowledges Department of Space (DoS), India for all the support at PRL, Ahmedabad and, the hospitality at IIT Kanpur during this work. S.\,K.\,G. acknowledges financial support from IIT Kanpur via the Initiation Grant (IITK/PHY/2022116). P.\,D. and S.\,K.\,G. also acknowledge Anusandhan National Research Foundation (ANRF) erstwhile Science and Engineering Research Board (SERB), India for financial support through the Start-up Research Grants: SRG/2022/001121 and SRG/2023/000934 respectively.


\appendix
\section{Lindblad formalism to describe dissipation due to the normal lead}
\label{Lindblad_formalism}

To investigate the impact of coupling a `lossy' metallic lead at the junction to an electron reservoir on the transport properties of a TIJJ, we model the system as an open quantum system interacting with its environment, following Ref.~\cite{Li2024}. The total Hamiltonian can be decomposed as $H=H_0+H_E+H_c$, where $H_0$ is the system Hamiltonian, $H_E$ describes the environment, and $H_c$ represents the coupling between the system and environment. The dissipative dynamics induced by this coupling term can be described by the Lindblad master equation~\cite{breuer2002,daley2014,roccati2022non,Manzano2020}:
\beq\label{Lindblad_eq}
    \frac{d\rho_0}{dt}=-i[H_0,\rho_0]+\sum_{m}\eta \left(\hat{\mathcal{L}}_m\rho_0 \hat{\mathcal{L}}^{\dagger}_m-\frac{1}{2}\{\hat{\mathcal{L}}_m \hat{\mathcal{L}}^{\dagger}_m,\rho_0\}\right).
\eeq
Here, $\rho_0$ is the reduced density matrix of the system, $\hat{\mathcal{L}}_m$ is the Lindblad (quantum jump) operator, and $\eta > 0$ is the dissipation rate. Rearranging this equation yields:
\begin{align}\label{Master equation}
   \frac{d\rho_0}{dt}=-i(H_{\textrm{eff}} \rho_0-\rho_0 H^{\dagger}_{\textrm{eff}})+\sum_{m} \eta \hat{\mathcal{L}}_m \rho_0 \hat{\mathcal{L}}^{\dagger}_m \;,
\end{align}
where, the effective Hamiltonian is 
\beq
H_{\text{eff}}=H_0-\frac{i}{2}\sum_{m} \eta \hat{\mathcal{L}}_m \hat{\mathcal{L}}^{\dagger}_m\;.
\eeq
The non-Hermitian term $-\frac{i}{2}\sum_{m} \eta \hat{\mathcal{L}}_m \hat{\mathcal{L}}^{\dagger}_m$ accounts for the `loss' due to coupling with the environment. By neglecting quantum jumps in the second term on the right-hand side of \equref{Master equation}, the system's dynamics can be effectively described by $H_{\text{eff}}$~\cite{Harrington2022,Li2024}. This motivates modeling the non-Hermiticity introduced by the `lossy' lead in the TIJJ by an imaginary potential term $-i V_2$ where $V_2 >0$ is real.

\begin{figure*}[!t]
\centering
\includegraphics[width=\textwidth]{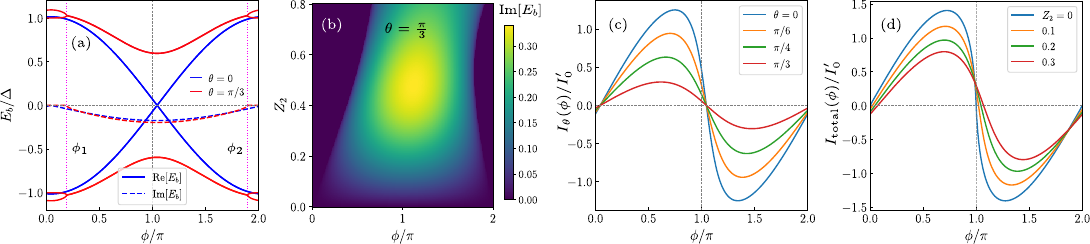}
\caption{\textbf{Characteristics of the NH-TIJJ in the strong-proximity limit ($\Gamma_0=\Delta$)}: (a) The real (solid lines) and imaginary (dashed lines) parts of the Andreev spectrum of the NH-TIJJ as a function of the phase difference $\phi$ plotted for different angles of incidence $\theta$. (b) Variation of the imaginary part of the spectrum as a function of $\phi$ and $Z_2$ for the incident angle $\theta=\pi/3$, the boundary of the dark purple region demonstrates the asymmetric gap-exit points in the spectrum. (c) The angle-resolved current $I_{\theta}(\phi)$ plotted as a function of $\phi$ for different angles of incidence. (d) The total supercurrent $I_{\mathrm{total}}(\phi)$ as a function of $\phi$ for different barrier parameters. The current is computed in units of $e\Delta/\hbar$ and we have chosen the real barrier parameter $Z_1=0.4$ and the imaginary barrier parameter $Z_2=0.2$}
\label{Appendix_panel}
\end{figure*}

\section{Wavefunctions and bound states for NH-TIJJs}
\label{wfn_NHTIJJ}
The wavefunctions for the two regions $x<0$ ($x_-$) and $x>0$ ($x_+$) of the NH-TIJJ, within the Andreev approximation and considering the limit of large Fermi energy ($\mu \gg \Tilde{\Delta}(\epsilon)$), obtained by solving \equref{BdG_eq} have the forms~\cite{Tkachov2013},
\begin{align}
   \psi_{k_y}(x_+)&=A\begin{bmatrix}
       1\\e^{i\theta} \non\\-e^{i\theta}a^+_R\\a^+_R \end{bmatrix} e^{ik_Fx\cos{(\theta)}-x/\xi} \\
        &+ B\begin{bmatrix}
        1\\-e^{-i\theta}\\e^{-i\theta}a^-_R\\a^-_R
        \end{bmatrix} e^{-ik_Fx\cos{(\theta)}-x/\xi} \;,\label{wf1}\\
   \psi_{k_y}(x_-)&=C\begin{bmatrix}
       1\\e^{i\theta}\\-e^{i\theta}a^-_L\\a^-_L
       \end{bmatrix} e^{ik_Fx\cos{(\theta)}+x/\xi} \non\\
        &+D\begin{bmatrix}
          1\\-e^{-i\theta}\\e^{-i\theta}a^+_L\\a^+_L
          \end{bmatrix} e^{-ik_Fx\cos{(\theta)}+x/\xi} \label{wf2},
\end{align}
where, $A$, $B$, $C$ and $D$ are constants, and $k_F$ is the Fermi wavevector. The Andreev reflection amplitudes for the electron-like ($+$) and hole-like ($-$) quasiparticles  are,
\begin{align}\label{andreev_coefficients}
    &a^{\pm}_{L,R}(\epsilon)=\frac{\Tilde{\Delta}(\epsilon)e^{-i\phi_{L,R}}}{\mathcal{E}(\epsilon)\pm i\sqrt{\Tilde{\Delta}^2(\epsilon)-\mathcal{E}^2(\epsilon)}}\;,\\
    &\cos{(\theta)}=\sqrt{1-\left(\frac{k_y}{k_F}\right)^2},~~ \mathcal{E}(\epsilon)=\epsilon+i\Gamma(\epsilon) \;.
\end{align}

Inserting \equsref{wf1}{wf2} in \equref{boundary}, we get a set of four equations for $A,B,C$ and $D$. For non-trivial solutions, the determinant of the coefficients of this system of equations must vanish, leading to the following equation for the ABS energy:
\bea\label{ABS_eq1}
    &T_1(a^{+}_L a^-_L + a^+_Ra^-_R) - T_2(a^{+}_L a^+_R + a^-_La^-_R) +\non\\
    & T_3(a^{-}_L a^+_R + a^+_La^-_R) -2i Z_1 Z_2 \cos^2{(\theta)}(a^{+}_L a^-_L - a^+_Ra^-_R) \non\\
    &+ 2Z_2 \cos{(\theta)}(a^{+}_L a^+_R - a^-_La^-_R)=0\;.
\eea
where $T_1=(1+Z^2_1-Z^2_2)\cos^2{(\theta)}$, $ T_2=Z^2_1+Z^2_2+\cos^2{(\theta)}$ and $T_3=\sin^2{(\theta)}(Z^2_1+Z^2_2)$. Substituting the coefficients from \equref{andreev_coefficients} in \equref{ABS_eq1} we arrive at \equref{gen_ABS_eq}. Now from \equref{full_gap_ABS}, denoting the R.H.S as $\mathcal{S}(\theta, \phi)$ and simplifying the L.H.S we obtain the following equation which can be rearranged to get \equref{ABS_eq},
\begin{equation}\label{B6}
    \left(\frac{\epsilon}{\Delta}+\frac{\epsilon}{\Gamma_0}\sqrt{1-\frac{\epsilon^2}{\Delta^2}} \right)^2=\mathcal{S}(\theta, \phi) \;.
\end{equation}
Substituting the form: $\left|\epsilon/\Delta \right|=c_1 \gamma+c_2 \gamma^2+c_3 \gamma^3$, in \equref{ABS_eq} and equating the coefficients of different powers in $\gamma$, we obtain
\bea
c_1 &=& -c_2=\mathcal{S}^{\frac{1}{2}}(\theta, \phi),\non\\
c_3&=&\mathcal{S}^{\frac{1}{2}}(\theta, \phi)+\frac{1}{2}\mathcal{S}^{\frac{3}{2}}(\theta, \phi) .
\eea

For the NH-TIJJ in the strong-proximity limit ($\Gamma_0=\Delta$), the Andreev spectrum is shown in \figref{Appendix_panel}(a) and (b) and bears similarities in qualitative behaviors with that of the tunneling limit (see \figref{NHTIJJ_complex}). This includes features like the spectrum bifurcation and asymmetry of the GEPs, note that for the strong-proximity limit the bifurcation happens around the parent superconducting gap $\Delta$ with one branch going above $\Delta$ and another below it, whereas in the tunneling limit the bifurcation occurs around the induced gap which is much smaller than $\Delta$. Similarly, the angle-resolved current (\figref{Appendix_panel}(c)) and the total current CPR (\figref{Appendix_panel}(d)) exhibit the anomalous behavior stemming from the asymmetry of the spectrum.

\section{Hamiltonian and bound states for an `ordinary' NH-JJ }
\label{NHJJ_wfns}
The BdG Hamiltonian for the `ordinary' NH-JJ (3DTI in \figref{fig:junction_schematic} is replaced by a 3D metal), in the same way as \equref{BdG_eq} in the main text, can be written as,
\beq
\label{NHJJ_BdG_Hamil}
H^{(0)}_{\text{BdG}}= 
\begin{bmatrix}
\hat{h}_0-\mu+U(x) & \Delta' e^{i\phi_{L,R}}\\
\Delta' e^{-i\phi_{L,R}} & -(\hat{h}_0-\mu+U^*(x))
\end{bmatrix}
\eeq
Here, $\hat{h}_0=-\frac{\hbar^2}{2m}(\partial^2_x+\partial^2_y)$ is the kinetic energy operator, $m$ is the mass of electron, $\mu$ is the chemical potential of the superconductor and $U(x)$ is the complex potential barrier. The induced pair potential $\Delta'$ is assumed to be energy independent. \\

The wavefunctions for the bound states are obtained by solving for the eigenstates of $H^{(0)}_{\text{BdG}}$ in \equref{NHJJ_BdG_Hamil}. Within the Andreev approximation ($k_h=k_e^*\simeq k_F$) and taking $p_y$ is a good quantum number, the general forms of the wavefunctions are given by: $\Psi(\textbf{r})=\sum_{k_y}\psi_{k_y}(x)e^{ik_yy}$, where,
\bea
\psi_{k_y}(x_-) &=&
                      A_1 e^{ik_hx\cos(\theta)}
                       \begin{pmatrix}
                            v\\u
                       \end{pmatrix}+
                    B_1 e^{-ik_ex \cos(\theta)}
                        \begin{pmatrix}
                            u\\v
                        \end{pmatrix} \;, \non\\
\!\!\psi_{k_y}(x_+) &=& C_1 e^{ik_ex \cos(\theta)}
                       \begin{pmatrix}
                        u e^{\frac{i\phi}{2}}\\ve^{-\frac{i\phi}{2}}
                       \end{pmatrix} \non\\ &+&
                    D_1 e^{-ik_hx \cos(\theta)}
                       \begin{pmatrix}
                        v e^{\frac{i\phi}{2}} \\u e^{-\frac{i\phi}{2}}
                       \end{pmatrix}\;, \non\\
u &=& \sqrt{\frac{1}{2}\Bigg(1 + \frac{i\sqrt{\Delta'^2-\epsilon^2}}{\epsilon}\Bigg)}\;, \non\\
v &=& \sqrt{\frac{1}{2}\Bigg(1 - \frac{i\sqrt{\Delta'^2-\epsilon^2}}{\epsilon}\Bigg)}\; \non,
\eea
where, $\cos(\theta) = \sqrt{1-(k_y/k_F)^2}$. The boundary conditions for the wavefunctions on either side of the junctions are,
\begin{align}
    &\Psi_-(0)=\Psi_+(0) \;, \non\\
    &\Psi'_+(0)-\Psi'_-(0)=-\frac{2m}{\hbar^2}(iV_1\tau_0+V_2\tau_z)\Psi(0) \;.
\label{NHBC}
\end{align}
The boundary conditions in \equref{NHBC} lead to four sets of equations for the amplitudes $A_1$, $B_1$, $C_1$ and $D_1$. For a non-trivial solution the determinant of the characteristic equation must be zero from which we obtain the secular equation to determine the ABS energy to be,
\begin{align}\label{NHJJ_ABS_eq}
    &\frac{\Delta'^2}{\epsilon^2} (Z^2 +\cos^2{(\theta)}\cos^2{(\phi/2)}) \non\\
    &-2i Z_2 \cos{(\theta)}\sqrt{\frac{\Delta'^2}{\epsilon^2}-1}-(Z^2+\cos^2{(\theta)})=0 \;.
\end{align}
Solving the above \equref{NHJJ_ABS_eq} for $y={\epsilon^2}/\Delta'^2$, we obtain the expression of ABS energy in \equref{ABS_energy_NHJJ}.\\

\bibliography{ref}

\end{document}